# Physics-informed deep learning links geodetic data and fault friction

Rikuto Fukushima[1], Masayuki Kano[2,3], Kazuro Hirahara[4,5], Makiko Ohtani[6]

[1] Department of Geophysics, Stanford University, Stanford, U.S.

[2] Disaster Prevention Research Institute, Kyoto University, Kyoto, Japan.

[3] Japan Agency for Marine-Earth Science and Technology

[4] RIKEN Center for Advanced Intelligence Project, Seika, Japan.

[5] Kagawa University, Kagawa, Japan.

[6] Graduate School of Science, Kyoto University, Kyoto, Japan.

Corresponding author: Rikuto Fukushima (rfukushima@stanford.edu)

## Abstract

Fault slip modeling, based on laboratory-derived friction laws, has significantly enhanced our understanding of fault mechanics. Agreement between model predictions and observations supports the hypothesis that observed slip diversity, including fast earthquakes and slow transient slips (Slow Slip Events; SSEs), originates from frictional heterogeneity. However, quantitative assessments of frictional heterogeneity from geodetic observations while fully incorporating fault mechanics are lacking due to the difficulties of high-dimensional optimization. In this study, we aim to address this gap using Physics-Informed Neural Networks (PINNs) to link frictional heterogeneity with geodetic observations. PINNs employ a neural network to represent the spatially variable frictional properties, making their estimation feasible. Targeting the 2010 Bungo SSE in southwest Japan, our estimation reveals heterogeneous friction coinciding with localized SSE nucleation in southwest Shikoku, and subsequent westward propagation. The calculated fault slip of SSE successfully reproduces the spatio-temporal pattern of observed surface displacements. This PINN-based inversion provides a mechanically consistent fault slip model validated through quantitative comparison with observations. Furthermore, we forecast the future fault slip evolution, demonstrating the importance of assimilating observations spanning multiple SSE cycles. Our results demonstrate the potential of PINN for advancing understanding of fault mechanics and enabling physics-based fault slip forecasting.

## Main Text

The construction of physics-based fault slip models is essential for understanding fault mechanics and for hazard risk assessment. Fault slip behavior has been modeled as a coupled process of rock elasticity and fault friction[1–3] by incorporating the laboratory-derived friction law: rate and state friction (RSF)[4–6]. As a consequence of different frictional properties on the fault, this framework reproduces various fault slip behaviors: interseismic stress accumulation, earthquake rupture, postseismic slip, and spontaneous slow slip event (SSE)[7–12]. The agreement between model predictions and observations supports the hypothesis that the observed fault slip diversity, ranging from fast earthquakes to SSEs, originates from frictional heterogeneity.

However, existing studies have often relied on limited comparisons between frictional models and observations. Conventional approaches first employ kinematic inversion to estimate slip evolution from geodetic and/or seismic observations and then compare this kinematically inverted slip with physics-based simulations. A major limitation of such kinematic inversions is the ad hoc regularization: kinematic source inversion requires the spatio-temporal regularization, such as Laplacian smoothing, to obtain stable solutions, but these constraints often produce over-smoothed slip distributions, potentially conflicting with complex rupture behavior[13,14].

Dynamic inversion offers a complementary approach to kinematic inversion, directly estimating physical parameters within a mechanistic model to match observations. This approach provides a promising path toward comprehensive validation of fault physics by bridging the gap between models and observations. Although existing forward simulations typically rely on dynamic parameters manually tuned by trial and error, data assimilation techniques offer a systematic approach for determining parameters. Data assimilation techniques have been applied to fault modeling, including postseismic deformation[15–17], coseismic rupture[18–22], and SSE[23–28]. This direction is crucial for fault slip forecasting, as assimilating interseismic geodetic observations into earthquake cycle models enables simulation of future fault slip evolution[29]. Frictional heterogeneity is a key control on fault slip, yet inverting for the spatial distribution of frictional parameters remains challenging. Spatially variable frictional parameters form high-dimensional vectors in discretized model, which makes the non-linear inversion more computationally intensive. Some previous studies have therefore imposed spatial constraints to reduce dimensionality[23,25], but such constraints preclude the fault slip complexities arising from heterogeneous friction.

Here, we employ Physics-Informed Neural Networks (PINNs), a recently developed deep-learning based assimilation method, for physics-based inversion of frictional heterogeneity. PINNs provide a flexible framework for data assimilation and have shown promise in static and dynamic elasticity problems[30–33], and laboratory experiments[34,35]. Crucially, PINNs represent spatially variable parameters as continuous functions via neural networks. Fukushima et al.[33] demonstrated through numerical experiments that PINNs can effectively estimate frictional heterogeneity from synthetic geodetic observation of SSEs. In this paper, we advance this research by assimilating real GNSS observations. Compared to fast earthquakes, SSEs have more moderate slip velocities and longer durations and is more feasible target of data assimilation[24]. The couple of models have been proposed as the possible mechanism of SSEs (e.g., brittle-ductile transition at a high slip velocity, pore-fluid pressure change due to porosity change[36], or frictional-viscous subduction zone model[37]), but the community do not have a common consensus for the mechanism of SSEs. This further motivates validation of physical models through dynamic inversion. In this study, we employ a simple SSE model governed by pure RSF as a first step, leaving more complex formulations for future work.

We target the SSE observed at the Bungo region of southwest Japan in 2010 (Figure 1a). Geodetic observations revealed that multiple SSEs with similar spatio-temporal evolution recur at regular intervals of approximately 6 yrs[38–40]. Several SSEs, including 2002-2004 and 2018-2019 events, comprise two subevents, and these complex yet repeating patterns suggest heterogeneous frictional properties of this region[39,40]. Kano et al.[25] (hereafter referred to as K24) conducted a Monte Carlo Markov chain-based estimation of frictional parameters using a simple circular SSE patch from the 2010 Bungo SSE. Although their results reproduced the long-term trend of observed surface displacement, they could not reproduce the rapid increase in the displacement, presumably due to an overly restrictive constraint assuming a circular patch with uniform frictional properties. To improve data fit and resolve the frictional heterogeneities causing the complexities of the 2010 Bungo SSE, we relax this constraint to estimate the spatial distribution of frictional parameters using the PINN-based method.

The main purpose of this research is to resolve frictional heterogeneities from the spatio-temporal evolution of the 2010 Bungo SSE using PINN-based inversion. Throughout its application to the 2010 Bungo SSE, we investigate the hypotheses that (i) PINN-based inversion can reveal spatially variable frictional properties consistent with surface observations; (ii) introduction of the fault mechanics provides the regularization in a more physically consistent way than ad hoc smoothing; (iii) this inversion can quantitatively assess the discrepancy between assumed physics model and observation; and (iv) inverted parameters provide the fault slip forecasting.

***Frictional heterogeneities with localized SSE initiation***

The PINN-based inversion successfully reproduced the spatio-temporal pattern of surface crustal deformation more closely than K24 (Figs. 1b and 2), resolving heterogeneous friction beneath southwest Shikoku (Fig. 3a). Velocity-weakening region with negative frictional parameter *a-b* extends beneath the Bungo Channel where the SSE occurs, surrounded by a velocity-strengthening region with positive *a-b* that inhibits SSE expansion (Fig. 3a). The most important feature is strong velocity-weakening properties with small characteristic slip distance *L* values in southwest Shikoku; smaller nucleation length[3] $R_c$ (Eq.(6), Method)) indicate stronger frictional instability (Fig. 3a), causing SSE nucleation in this region (Fig. 3b). This is consistent with observed surface displacement, which shows slip initiation from southwest Shikoku during 2009.7 to 2010.1, followed by westward slip expansion to the Bungo Channel with large slip during 2010.1 to 2010.7 (Fig. 2). GNSS station 0085 (D at Fig.1) in southwest Shikoku recorded gradual displacement increase from 2009.7 (Fig. 1b), reflecting SSE growth beneath southwest

Shikoku. In contrast, displacement in Kyushu showed no major signal during 2009.7-2010.1 but increased abruptly from 2010.1 (Fig. 1b and Fig. 2), owing to westward propagation of large slips from southwest Shikoku. Unlike the PINN-based inversion with spatially variable friction, the K24 model with homogeneous friction produces uniform gradual nucleation over the entire SSE region (Extended Data Fig. 1b) and cannot explain the abrupt slip velocity increase (Fig.1b). These results demonstrate the importance of spatially heterogeneous friction to describe the spatio-temporal evolutions of SSE.

The frictional heterogeneity in southwest Shikoku is consistent with previous kinematic analysis. Aseismic slip initiation from southwest Shikoku followed by extension to the Bungo Channel, has been reported for multiple events, including 1997, 2003, and 2010 SSEs[38]. These similar SSE behaviors have been interpreted to reflect frictional heterogeneity, although no previous research has directly linked this interpretation to RSF parameter, which is a fundamental limitation of kinematic inversion. In contrast, our proposed PINN-based inversion directly links frictional properties with geodetic observation through RSF, providing a physics-based interpretation of observed slip events.

As RSF parameters exhibit strong trade-offs[41], we should carefully assess which frictional parameters are robustly estimated. We performed the same inversion but removed the inequality constraint for frictional parameters, which were imposed to ensure consistency with laboratory experiments (Method). Without this constraint, inversion resolves different frictional parameters on the creeping region but consistently estimates the small nucleation length on the southwest Shikoku (Extended Data Fig. 10). This leads us to conclude that at least heterogeneity for $R_c$ in the SSE region should be a robust result constrained from geodetic data. To further elucidate parameter trade-offs, a Bayesian

PINN framework is a promising approach for quantifying the posterior distribution of estimated parameters[42–44].

### *Physics-based regularization of inversion*

A key advantage of inversion incorporating fault mechanics is physics-based regularization. The friction law provides the physics-based constraints for the temporal slip evolution and restricts the admissible solution space, decreasing the non-uniqueness compared with kinematic inversion. Instead of ad hoc regularization such as Laplacian smoothing, the PINN formulation employs the fault mechanics residual assuming elasticity and RSF ($J_{\mathrm{ode}}$; Eq. 9, Method) and the prior on the fault strength assuming weak locking during the interseismic phase ($J_{\mathrm{ini}}$; Eq.12, Method). Those regularize fault slip in a physics-consistent manner, providing a complementary approach to traditional kinematic inversion. Here, we compare our estimation result with traditional kinematic inversion in terms of physics-based regularization.

In our estimation, during 2010.3-2010.6, aseismic slip expanded to the Bungo Channel concurrent with slip deacceleration in southwest Shikoku (Fig. 3b). Because accumulated stress in southwest Shikoku had been released before this period, spatial slip separation becomes clear at later stage of SSE (for example, 2010.5-2010.6). This spatio-temporal evolution is not clear in previous kinematic inversion[39], presumably because spatio-temporal smoothing renders slip separation more ambiguous. In contrast, PINN-based inversion naturally imposes spatio-temporal regularization through RSF-based fault mechanics[33], yielding a physically consistent slip model. This result demonstrates the potential of PINN to provide physics-based regularization in fault slip estimation.

While our PINN-based method does not impose explicit smoothing, there is a possibility that the neural network architecture provides the implicit smoothing to the solution. Inversion for the spatial distribution of RSF parameter can result in short-wavelength artifacts at a local minimum when appropriate regularization is absent[22]. Fully-connected neural networks are biased to be long wavelength function (spectral bias)[45], and this could work as the implicit smoothing that suppresses short wavelength artifacts. Clarifying how such implicit smoothing originates from neural network architecture is an important topic for future work.

***Towards the understanding of the SSE mechanism***

Quantifying discrepancies between model predictions and observations is crucial for validating physical mechanism. Although our model broadly explains the long-term deformation time series, we could identify a small difference in SSE nucleation timing around 2010.1 (Fig. 1b, Extended Data Fig. 2). Our PINN model initiates slip slightly earlier, whereas observed displacement shows later and more rapid increase. This results in slip overestimation during 2010.0-2010.1 and underestimation during 2010.1-2010.3 (Extended Data Fig. 3). This misfit may indicate the need to consider alternative physical mechanisms (for example, postseismic creep following slow ruptures[46], brittle-ductile transition at a high slip velocity[47–49], pore-fluid pressure change due to porosity change[36], or frictional-viscous subduction zone model[37]), although further investigation is required (for example, incorporating more complex fault geometry such as the 3-D plate interface or elastic structure heterogeneity). The physical mechanism of SSEs could be discussed in the future by quantifying the model discrepancy of different SSE model. Notably, such detailed evaluation of physical mechanisms becomes possible only after achieving quantitative inversion for frictional heterogeneity that substantially improves data fit.

PINNs enable significant progress by facilitating high-dimensional inversion that resolves the spatial structure.

### *Future direction towards fault slip forecasting*

Figure 4 shows slip prediction results based on the estimated parameters. Following the 2010 event (moment magnitude (Mw) 6.8), the model predicts SSE initiation from southwest Shikoku in 2014, lasting approximately 1 year with comparable moment magnitude (Mw 6.7). The predicted SSE characteristics– moment magnitude, recurrence period, and event duration, are consistent with typical Bungo SSE values. Kinematic analysis of GNSS data[50] detected aseismic slip in southwest Shikoku in 2015, comparable in timing to our prediction, although the observed slip magnitude was substantially smaller.

The predicted 2014 SSE is more extreme, with higher maximum slip velocity (approximately 15 $V_{pl}$ where $V_{pl}$ is plate convergence rate). The spatial slip distribution differs from the 2010 event, with larger slip in southwest Shikoku and smaller slip in the Bungo Channel, producing smaller surface displacement in Kyushu (Extended Data Fig. 4). This indicates that the estimated fault state following the 2010 event had not yet reached limit cycle, which is a phase of perfectly periodical behavior. In the limit cycle, southwest Shikoku is more strongly locked than before the 2010 event, producing more extreme SSEs with maximum slip velocity of ~162 $V_{pl}$ (Extended Data Fig. 5b). Notably, the limit-cycle moment magnitude (Mw 6.8) and recurrence interval (~6 years) remain consistent with typical Bungo SSE (Extended Data Fig. 6).

These results demonstrate the importance of simultaneously assimilating fault state and frictional parameters to account for the interevent fault locking. We first remove the linear interseismic trend from the data (Method), and therefore the locking information inferred from interseismic deformation is lacking in our inversion. Interseismic locking data should infer the interseismic creeping and locked asperity erosion[51], contributing to constrain frictional properties (i.e., *a*-*b* at the velocity-strengthening region[29]). The consideration of interseismic phase is therefore a crucial future direction, enabling more reliable fault slip forecasting. Other import direction is multiple-event assimilation: continuous geodetic data enable tracking of fault state through both locking and slipping periods, allowing simultaneous estimation of initial stress and frictional parameters without prior constraints on initial state variable $\theta_{\mathrm{ini}}$ (Method). Instead of using multiple event data, we could simply impose the additional constraint that the fault slip behaves cyclic to obtain limit cycle solution[26].

Our work quantitatively reveals frictional heterogeneity underlying the spatio-temporal evolution of SSEs. Physics-based inversion constrains spatio-temporal slip pattern through mechanically consistent regularization, reveals small discrepancies between physical models and observations during nucleation, and gains insight into fault slip forecastability. PINN-based fault slip forecasting is further explored in the successive study[52]. Further investigation incorporating diverse observations or physical models will advance understanding of fault mechanics and enable physics-based fault slip forecasting.

## Methods

### *Framework of fault physics*

Our simulation framework is based on RSF[4,5]. To model the Bungo SSE, we employed a 120 × 100 km rectangular fault for strike and dip direction, subducting with a dip angle of 15°, referring to the previous data assimilation research targeting the same region[23,25,26,33]. We divided the fault into 2 × 2 km subfaults. In the RSF framework, the fault frictional stress $\tau$ is the function of fault slip velocity $V$ and state variable $\theta$:

$$\tau = \tau * + \sigma a \log\left(\frac{V}{V*}\right) + \sigma b \log\left(\frac{\theta}{\theta*}\right), \#(1)$$

where σ is the normal stress, $a$, $b$, and $L$ are frictional parameters, and $\tau*$ denotes the shear stress at the reference velocity $V^*$ and reference state variable $\theta^*$. We assume the uniform normal stress as $\sigma$ = 100 MPa. Since $\sigma$ and frictional parameters ($a$ and $b$) cannot be distinguished in this RSF formulation, we estimate $\sigma a$ and $\sigma b$ in inversions, and convert to $a$ and $b$ by assuming this normal stress. The temporal evolution of state variable $\theta$ can be described in several ways, and we used the aging law[5], where $\theta$ obeys the following equations:

$$\frac{d\theta}{dt} = 1 - \frac{V\theta}{L}. \#(2)$$

Here we introduce the variable $\Omega = V\theta / L$, which represents the ratio of weakening to healing rates[53]. $\Omega = 1$ results in a steady state, where the state variables do not change due to the balance of healing and weakening, and then stress only depends on the slip velocity $V$;

$$\tau_{ss} = \tau * + \sigma(a-b)\log\left(\frac{V}{V*}\right). \#(3)$$

With the steady state stress $\tau_{ss}$, the fault stress can also be written as $\tau = \tau_{ss} + \sigma b \log \Omega$. Since the dynamic stress change due to the wave propagation does not matter in the SSE, the elastic stress can be approximated in the form of a quasi-dynamic framework with radiation damping[54]:

$$\frac{d\tau_i}{dt} = \sum_j K_{ij}(V_j - V_{pl}) - \frac{G}{2c}\frac{dV_i}{dt}, \#(4)$$

where $V_{pl}$, $G$, and $c$ are the fault loading velocity, shear modulus, and shear wave velocity, respectively. $K_{ij}$ is a shear stress change at subfault $i$ due to the unit slip at subfault $j$, derived from the analytical solution of static elasticity in a linear isotropic elastic homogeneous half-space medium[55]. In our estimation, we fixed $V_{pl}$ = 6.5 cm/yr[56], $G$ = 40 GPa, and c = 3km $s^{-1}$. Combining Eqs. (1), (2) and (4), we get

$$\frac{dV_i}{dt} = \left(\frac{a_i\sigma}{V_i} + \frac{G}{2c}\right)^{-1} \left(\sum_j K_{ij}\left(V_j - V_{pl}\right) - b_i\sigma\left(\frac{1}{\theta_i} - \frac{V_i}{L_i}\right)\right). \quad (5)$$

RSF theory predicts that the system instability is determined by the slip patch length scale $R$ and nucleation length $R_c$ defined as[57]

$$R_c = \frac{\pi}{4}\frac{GbL}{\sigma(b-a)^2}. \#(6)$$

$R > R_c$ exhibits instability and causes the unstable slip earthquake, whereas $R << R_c$ causes the stable slip. Weak instability occurs with a limited range of parameters when $R$ is smaller but not so much as $R_c$, reproducing the spontaneous slow slip comparable with observed SSEs[10,23].

***GNSS Observation data***

We used the same dataset as K24. The surface displacement caused by our target SSEs occurred at the Bungo Channel has been observed by the GNSS Earth Observation Network System (GEONET) maintained by the Geospatial Information Authority of Japan[58]. We used daily time series of 86 GEONET stations around the Bungo Channel that were used in K24 (see figure S1 of K24). The data period is from July 2006 to June 2012, which is one year longer than K24, and the number of observation points $N_t$ is 2192. We estimate the linear trend at each observation station from the period July 2006 – June 2008, when SSE did not occur, and use the detrended time series as observation data to subtract this trend.

The residuals between the PINN output and the observation data are defined as;

$$J_{data} = \frac{1}{N_{data}N_t}\sum_{i=1}^{N_t} (\mathrm{H}\boldsymbol{s}_{\mathrm{NN}}(t_i) - \boldsymbol{d}_i)^{\mathrm{T}}\mathrm{C}^{-1}(\mathrm{H}\boldsymbol{s}_{NN}(t_i) - \boldsymbol{d}_i). \#(7)$$

where $\boldsymbol{d}_{\mathrm{i}}$ is the vector including three components of surface displacement at each station at time $t_i$, H is the observation matrix that converts fault slip to surface displacement at each station, calculated by analytical solution of static elasticity in a linear isotropic elastic homogeneous half-space medium[55], and C is the covariance matrix. $s_{\mathrm{NN}}$ is the estimated detrended fault slip obtained by the time integration of $V_{\mathrm{NN}}$. For computational efficiency, we use a first-order Euler method with $\Delta t$ = 1 day, which can be justified considering the slow deformation time scale of SSEs.

For simplicity, we subtract the linear trend estimated from the MAP solution of K24, assuming that the K24 model is a good approximation of interseismic locking during that period. Note that the estimated linear trend from actual data differs with that of K24 model or what we calculate in PINNs. This is because PINN/K24 model assume the steady slipping at $V_{\text{pl}}$ outside the fault (Fig. 1a) but shallower part not modeled by our fault is actually locking and contributes to the linear trend at each station. More exploration of linear trend subtraction is expected in the future, since the linear trend should include the information on interseismic stress accumulation. It should be valuable for the data assimilation of the whole SSEs cycle, and further, the forecasting of SSEs.

***Physics-Informed Neural Network setting***

Our design of Physics-Informed Neural Network is based on Fukushima et al.[33], where numerical experiments have been conducted to successfully estimate spatially variable frictional properties. We construct two fully-connected neural networks (NN); NN A represents the spatio-temporal evolution of simulation variables $V$ and $\theta$, and NN B represents the spatial distribution of the frictional parameters $a$, $a$-$b$, and $L$ (Extended Data Fig. 7a). We define $V_{\text{NN}}$, $\theta_{\text{NN}}$, $a_{\text{NN}}$, $a$-$b_{\text{NN}}$, and $L_{\text{NN}}$ as $V_{\text{NN}} = V^{*}\exp(\text{f}^{\text{A}}{}_{1})$, $\theta_{NN} = \theta^{*}\exp(\text{f}^{\text{A}}{}_{2})$, $a_{\text{NN}} = 10^{-3} \times \text{f}^{\text{B}}{}_{1}$, $a$-$b_{\text{NN}} = 10^{-3} \times \text{f}^{\text{B}}{}_{2}$, and $L_{\text{NN}} = 10^{-3} \times \text{f}^{\text{B}}{}_{3}$, where $\text{f}^{\text{A}}{}_{\text{i}}$ and $\text{f}^{\text{B}}{}_{\text{i}}$ is the $i$ th output of NN A and B. NN A consists of an input layer with three nodes, eight hidden layers with 20 nodes, and an output layer with two nodes, and NN B's architecture is the same as NN A except for two nodes of the input layer and three nodes of the output layer. The total number of neural network parameters is 6125. We choose the L-BFGS method[59] to optimize the neural network parameters. Since no robust theory has been established to determine the appropriate NN architecture or optimization method, we employ the same architecture and optimization used in the previous PINN-based numerical experiments[33].

PINN-based inversion has been done by optimizing the NN parameter based on the loss function, which incorporates physics, observation, and other prior information. We define the loss function $J_{\text{total}}$ as:

$$J_{total} = J_{ode} + w_{data} J_{data} + w_{ini} J_{ini}, \#(8)$$

where $J_{data}$, $J_{ode}$, and $J_{ini}$ represent data misfit (Eq. 7), the residuals of governing equations (Eq. 9), and the regularization term for initial condition (Eq. 12), respectively. This loss function can be seen as the negative logarithm of the Gaussian probability density function, and weight $w_{ini}$ and $w_{data}$ can be written as $w_{ini} = \sigma_{ode}^2 / \sigma_{ini}^2$ and $w_{data} = \sigma_{ode}^2 / \sigma_{data}^2$ by using the variance of each loss function: $\sigma_{ini}^2$, $\sigma_{ode}^2$, and $\sigma_{data}^2$ (Ref. Appendix A of Fukushima et al.[33]). $J_{ode}$ is defined to introduce physics-based differential equations to PINNs through the normalized residuals of governing equations;

$$J_{ode} = \frac{t^{*2}}{N_{ct}\, N_{cxy}} \sum_{i=1}^{N_{ct}} \sum_{j=1}^{N_{cxy}} \left( \underline{r_V}(t_i, x_j, y_j)^2 + \underline{r_\theta}(t_i, x_j, y_j)^2 \right), \#(9)$$

$$\underline{r_V} = \frac{1}{V_{NN}} \left( \frac{dV_{NN}}{dt} - f(V_{NN}, \theta_{NN}) \right), \#(10)$$

$$\underline{r_\theta} = \frac{1}{\theta_{NN}} \left( \frac{d\theta_{NN}}{dt} - g(V_{NN}, \theta_{NN}) \right), \#(11)$$

where the coupled equations of elasticity and friction (Eqs. (2) and (5)) are represented as $dV/dt = f(V, \theta)$ and $d\theta/dt = g(V, \theta)$. This loss is evaluated at the discrete points, which are called collocation points. Here, collocation points are set as equally distributed $N_{cxy}$ = 3,000 points for space and $N_{ct}$ = 219 for time. While the intervals of collocation points ($\Delta t$ = 10 days) is longer than the GNSS sampling interval, the continuity of neural networks provides the appropriate interpolation[32,33]. $J_{ode}$ is non-dimensionalized by the characteristic time of the system[60]: $t^* = L_{prior} / V_{pl}$, where $L_{prior}$ is the prior information for $L$ introduced in the following regularization section. The definition of $J_{ode}$ is identical to that used in the previous research[32,33].

### *Prior information for regularizing inversion*

To obtain a reasonable solution from the high-dimensional inverse problem, we employ the following strategies to regularize the inversion based on prior information. The inverse problem here is to find the spatial distribution of frictional parameters and the initial state variables based on the surface displacement. Previous research[26,33] has pointed out the difficulty of simultaneously inverting the initial state variable $\theta$ and the frictional parameters from a single cycle observation. Following the method used in Fukushima et al.[33], we introduce the regularization term for the state variable defined as:

$$J_{ini} = \frac{1}{N_{cxy}} \sum_{i=1}^{N_{cxy}} \left( \log \frac{V_{NN}(0, x_i, y_i)\theta_{NN}(0, x_i, y_i)}{L_{prior}} \right)^2 , \#(12)$$

where $L_{\text{prior}}$ is the prior frictional parameters. If $L = L_{\text{prior}}$, this loss function turns into $\| \log \Omega_{\text{ini}} \|^2 = \| (\tau_{\text{ini}} - \tau_{\text{ss}} (V_{\text{ini}})) / \sigma b \|^2$ and can be interpreted to assume that the initial fault stress $\tau_{\text{ini}}$ is not far away from the steady state. From the perspective of maximum likelihood estimation, this loss function introduces the prior probability function: $p(\theta_{\text{ini}}) = A \exp (- \| \log \Omega_{\text{ini}} \|^2 / 2\sigma_{\text{ini}}^2)$, and its weight $w_{\text{ini}}$ should be determined by the variance $\sigma_{\text{ini}}^2$, which represents the plausible value of $\log \Omega_{\text{ini}}$. We choose $L_{\text{prior}} = 4.0\times10^{-2}$ [m] based on the previous research targeting Bungo SSE[23]. Without this prior, the PINN inversion converges to poor local minima in which $J_{\text{ode}}$ remains large, and the resulting estimates are physically implausible. In practice, providing a reasonable prior for $L$ helps guide the inversion toward plausible solutions. Note that this uniform $L_{prior}$ might also introduce a certain smoothing effect.

***Hyperparameter choice***

While the maximum likelihood estimation provides the expression of loss weights with variance $\sigma_{\text{ini}}^2$, $\sigma_{\text{ode}}^2$, and $\sigma_{\text{data}}^2$, the choice of weight parameters is a challenging problem due to the complicated shape of the loss function in the high-dimensional neural network parameter space. In this paper, instead of providing a solid procedure to tune the hyperparameters, we just seek the weights by trial and error and find the appropriate parameters where the PINN output satisfies both the governing equations and data fit.

We finally set the weights as $w_{\mathrm{ini}} = 10^{-4}$ and $w_{\mathrm{data}} = 10^{-3}$. Extended Data Fig. 7b shows the learning curve with these parameters, and the loss function converges to $J_{\mathrm{ini}} = 1.25\times10^{-1}$, $J_{\mathrm{ode}} = 2.59\times10^{-5}$, and $J_{\mathrm{data}} = 2.57$. To check whether PINN's output obeys the governing equations, we solve the governing equations using the Runge-Kutta method[61], with the PINN estimated $a$, $a$-$b$, $L$, $V_{\mathrm{ini}}$, and $\theta_{\mathrm{ini}}$. The solution by the traditional solver reproduces the PINN's output (Extended Data Fig. 8), proving that $J_{\mathrm{ode}}$ becomes small enough values to satisfy the governing equations. Since $J_{\mathrm{ode}}$ and $J_{\mathrm{data}}$ seem to converge well, we finally adopt the PINN estimation with these hyperparameters.

***Transfer Learning***

To help the convergence of neural network parameters, we utilize transfer learning. Before applying the real data, we conducted the PINN-based inversion for the frictional parameter from the synthetic GNSS data and used that trained model as the initial neural parameter for real data assimilation as follows. We have generated the synthetic observation data by numerically simulating the fault slip with the frictional parameters estimated in K24 during the same periods of real data, and adding the Gaussian noise based on the real covariance matrix. The initial condition for this simulation is the same as K24 and assumed to be known during the PINN estimation by defining $J_{\mathrm{ini}} = ||V - V_{\mathrm{ini}}||^2 + ||\theta - \theta_{\mathrm{ini}}||^2$. The weights are set to $w_{\mathrm{ini}} = 1$ and $w_{\mathrm{data}} = 10^{-2}$ during pre-optimization. Extended Data Fig. 9 indicates the initial neural network parameters before real data assimilation, which resulted from the pre-optimization with synthetic data. Reflecting the K24 model, frictional parameters have a circular velocity-weakening patch, and surface displacement fits to K24 instead of real observation data. Starting from this model, frictional parameters are optimized to fit the real data during the main optimization.

***Constraints for frictional parameters***

Taking the consistency of the typical frictional parameters measured by laboratory experiments into account, we limit the possible range of frictional parameters for $a$ and $b$. We constrain the frictional parameters to satisfy $0 \leq a \leq 10^{-2}$ and $0 \leq b \leq 10^{-2}$, which are plausible values suggested from experiments, by cutting off the output of NN, as $a_{\text{est}} = \min(10^{-2}, a_{\text{NN}})$.

It should be noted that these constraints are not necessarily required to reproduce the observation data. Extended Data Fig. 10 shows the result of PINN estimation without frictional constraints. The estimated parameters have a small nucleation length in southwest Shikoku, which is the same characteristic as the PINN estimation with constraints (Fig. 3a), and can reproduce similar surface displacements with PINN estimation with frictional constraints. The large difference lies in the frictional parameters outside the patch: $a$ and $b$ become large values ($a \sim 7\times10^{-2}$, $b \sim 7.2\times10^{-2}$), which are far away from the estimated values in laboratory experiments. These results suggested that the frictional parameters outside the SSEs patch cannot be constrained only from the surface displacements, and hence, the prior information from laboratory experiments is useful to constrain values in a plausible range.

**Data Availability**

The GNSS data from GEONET[58] were provided by Professor Takuya Nishimura of the Disaster Prevention Research Institute, Kyoto University[62]. The coordinate time series themselves can be obtained via Zenodo repository[63], https://doi.org/10.5281/zenodo.14242258. The Python code for PINN-based inversion can be obtained via Zenodo repository, https://doi.org/10.5281/zenodo.14977672[64].

## Methods References

**Acknowledgements**

We thank Yusuke Tanaka for his support in analyzing the data and Daisuke Sato for his valuable comments on the PINN estimation results. This study was supported by the MEXT STAR-E Project [Grant JPJ010217], MEXT STAR-E NEXT Project [Grant JPJ013735], ERI JURP 2025-B-01 in Earthquake Research Institute, the University of Tokyo, and the JSPS KAKENHI [Grants 23H00466, 24K02951, 24H01019, 25K01084, 23K03552].

**Author Contributions**

R.F. and M.K. designed the study. R.F. carried out the inversion and prepared the manuscript. M.K, K.H., and M.O. advised the project. All authors discussed the results in the article.

**Competing Interests**

The authors declare no competing interests.

**Materials & Correspondence**

Correspondence and requests for materials should be addressed to R.F.

**Figure**

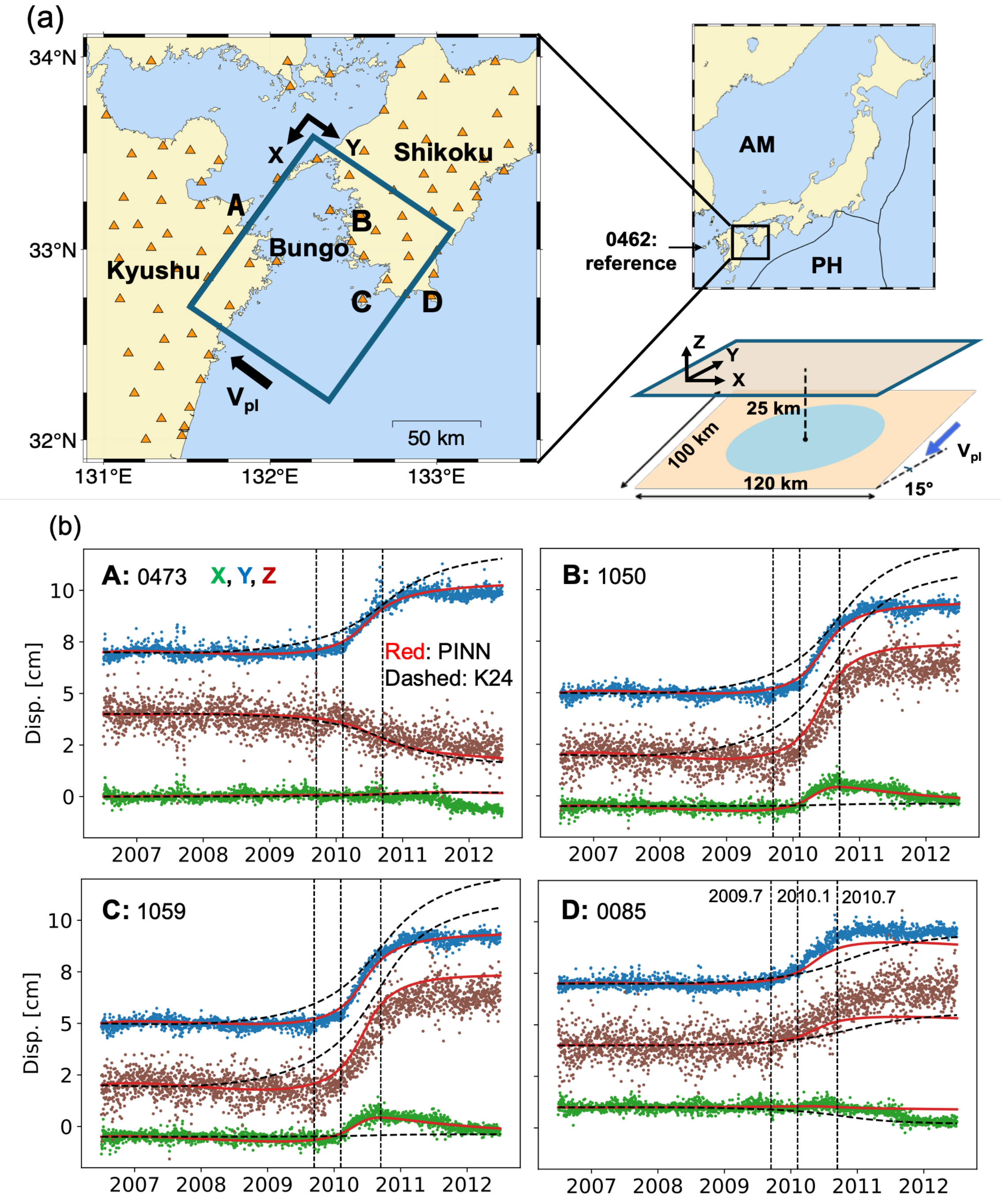

**Figure 1.** Target area and GNSS time series. (a) The geometry of southwest Japan. The orange triangles represent the 86 GEONET stations used in this research. The blue rectangle represents the assumed fault to model the observed SSE projected to the surface. The rectangular fault is 120 km along strike and 100 km along dip. The dip angle is 15°

and the depth of fault center is 25 km. Loading velocity is 6.5 cm/yr. AM and PH represent the Amurian and Philippine Sea plates, respectively. (b) Time series of surface displacement observed at 4 stations, A-D. Green, blue, and brown dots represent the observation data in X (trench-parallel), Y (trench-perpendicular), and Z (vertical) components, respectively. Black and red lines indicate the calculated displacement at K24 and PINN-based inversion, respectively. Three vertical dashed lines indicate the times of 2009.7, 2010.1, and 2010.7, respectively.

GNSS = Global Navigation Satellite System

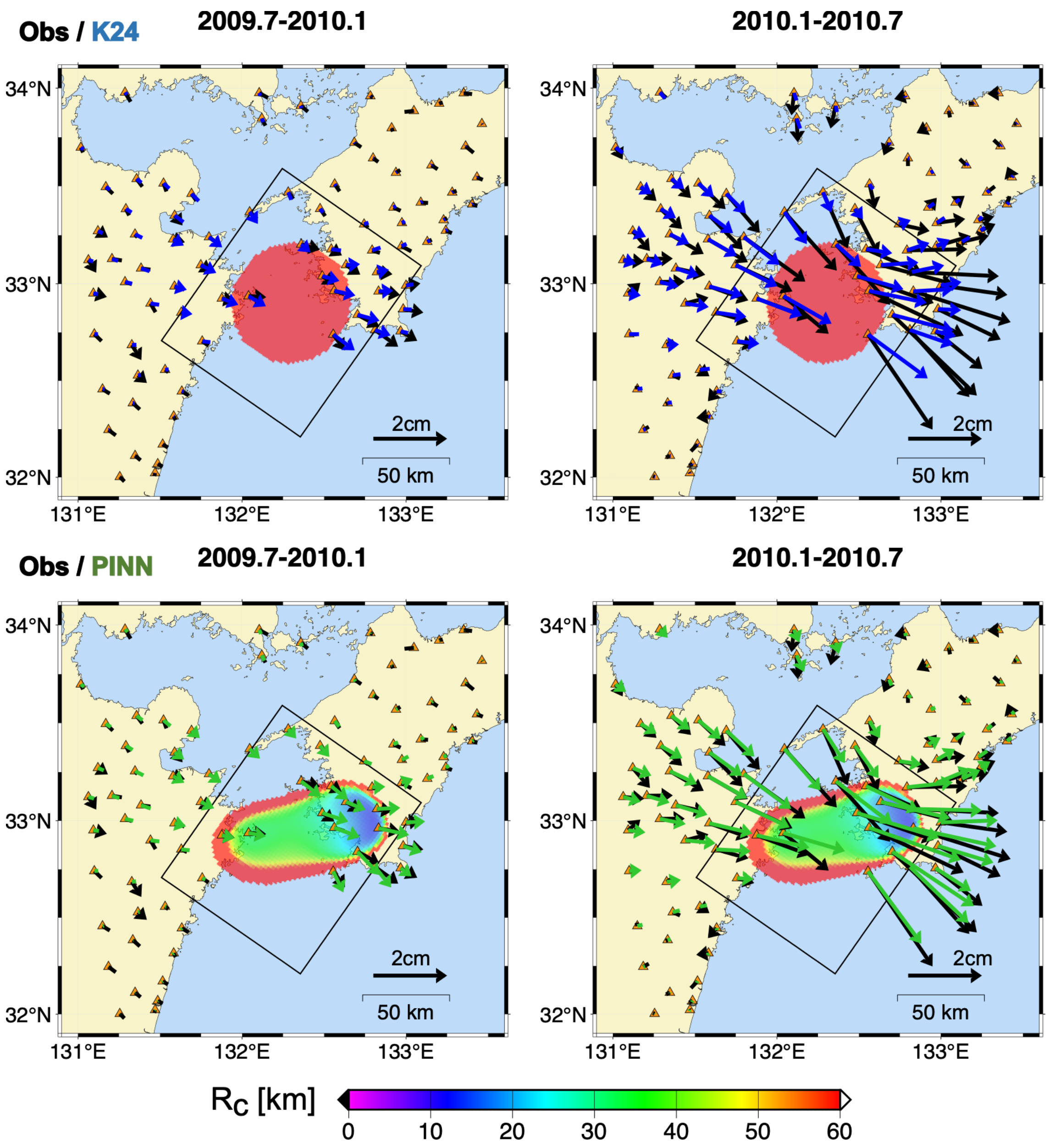

**Figure 2.** Displacement field during 2009.7-2010.1 (left panel) and 2010.1-2010.7 (right panel). Black, blue, and green arrows represent the displacement of observation, the K24 model, and the PINN-based model, respectively. The nucleation length $R_c$ on the velocity weakening region, estimated in the K24 model and PINN-based model, are plotted on the fault.

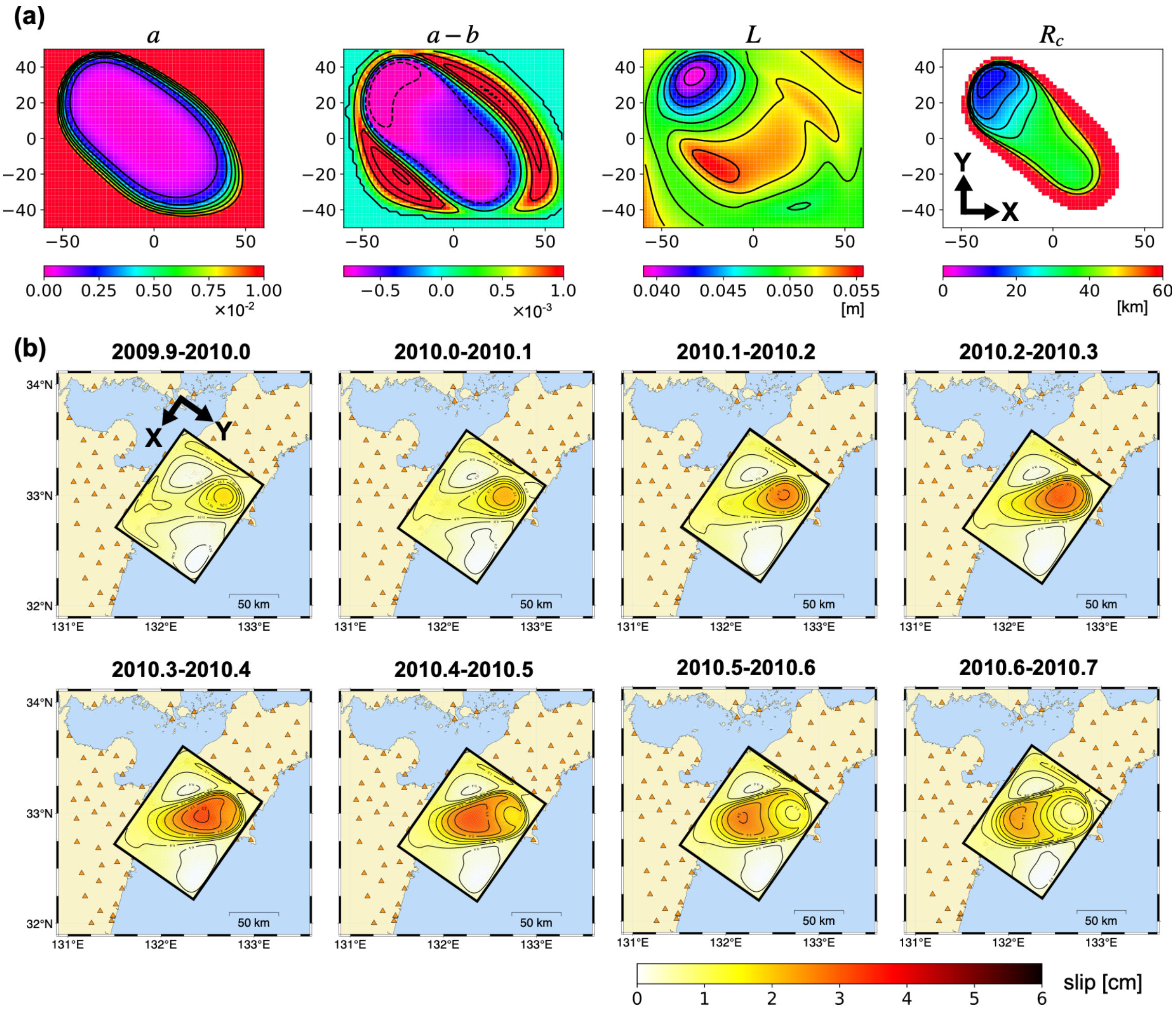

**Figure 3.** Results of the PINN-based estimation. (a) Spatial distribution of estimated frictional parameters for *a*, *a-b*, and *L*. $R_c$ shows the nucleation length (Eq. 6) at the velocity weakening region calculated from the estimated parameters. Each contour corresponds to $R_c$ = 15, 20, 25, 30, 40 and 50 km, respectively. (b) Estimated fault slip evolution during 2009.9-2010.7. The fault slip nucleated from southwest Shikoku, followed by its westward propagation.

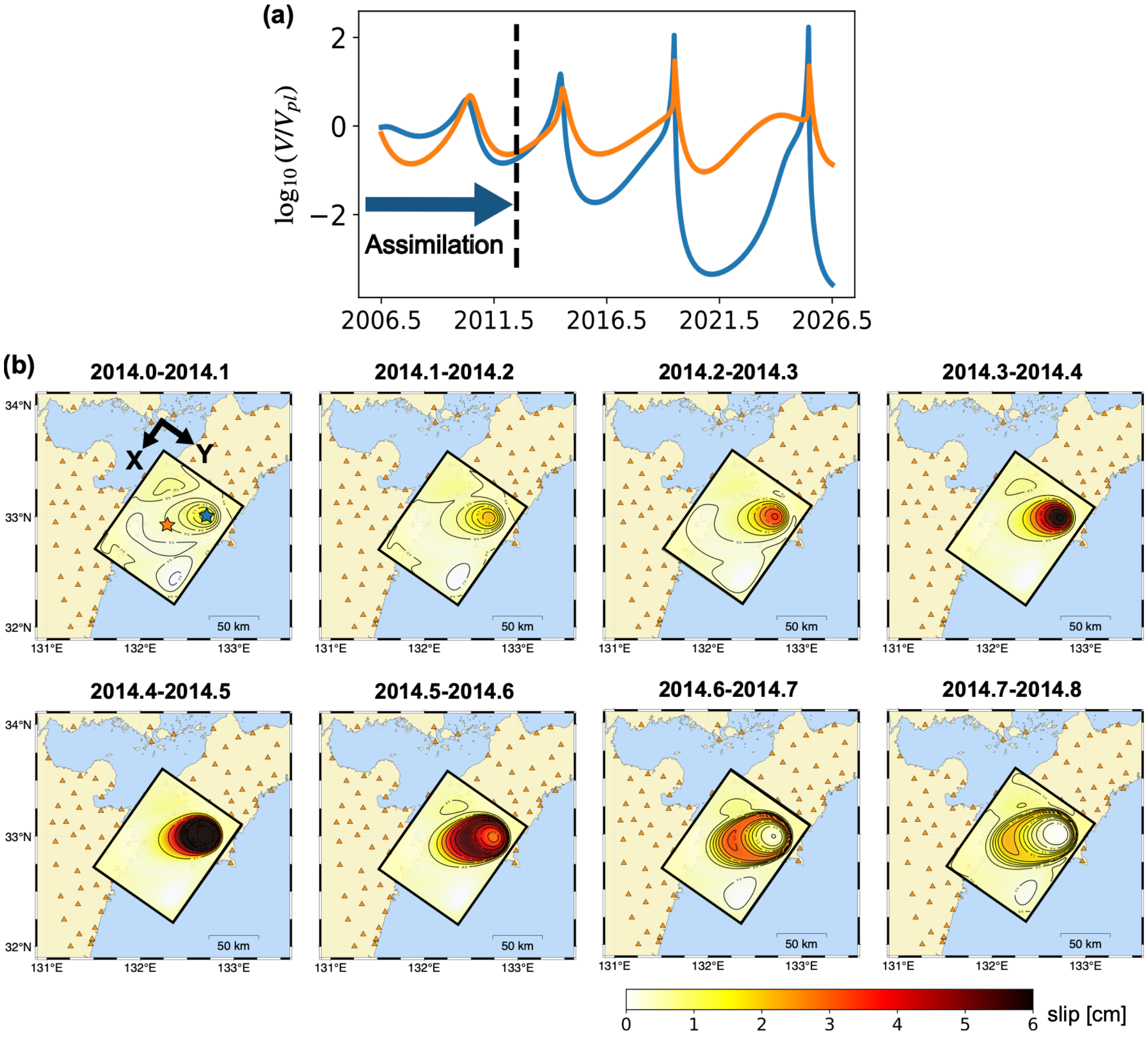

**Figure 4.** Results of fault slip prediction with PINN estimation results. (a) Fault slip evolution until 2026.5. Blue and orange lines represent the fault slip velocity at the lowest $R_c$ and center of the fault, which are indicated as blue and orange stars in (b), respectively. (b) Predicted fault slip during 2014.0-2014.8.

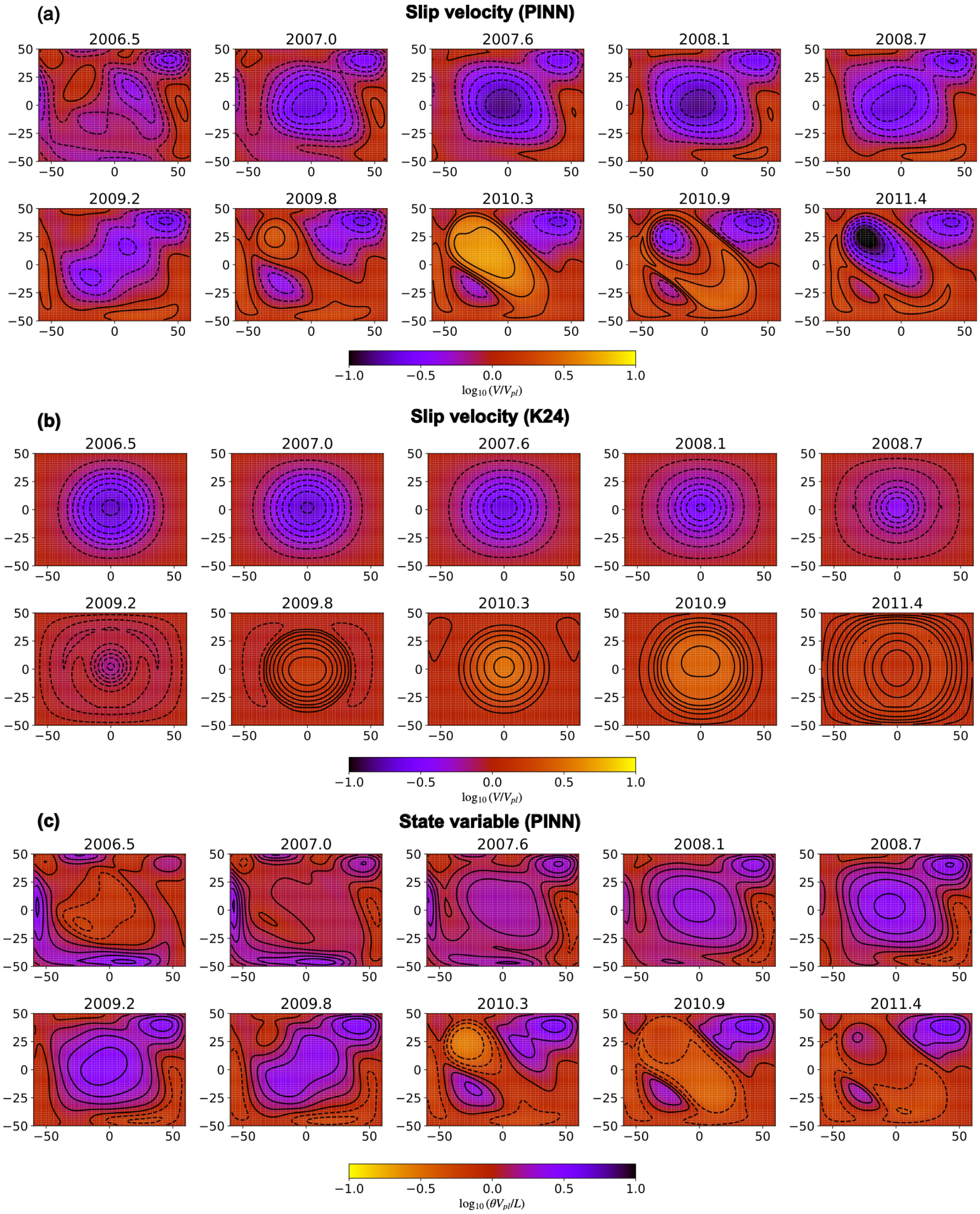

**Extended Data Figure 1.** (a) Estimated fault slip velocity during 2006.5-2011.4 by PINN-based inversion (b) Same with (a) but for K24 (c) Estimated state variable during 2006.5-2011.4 by PINN-based inversion

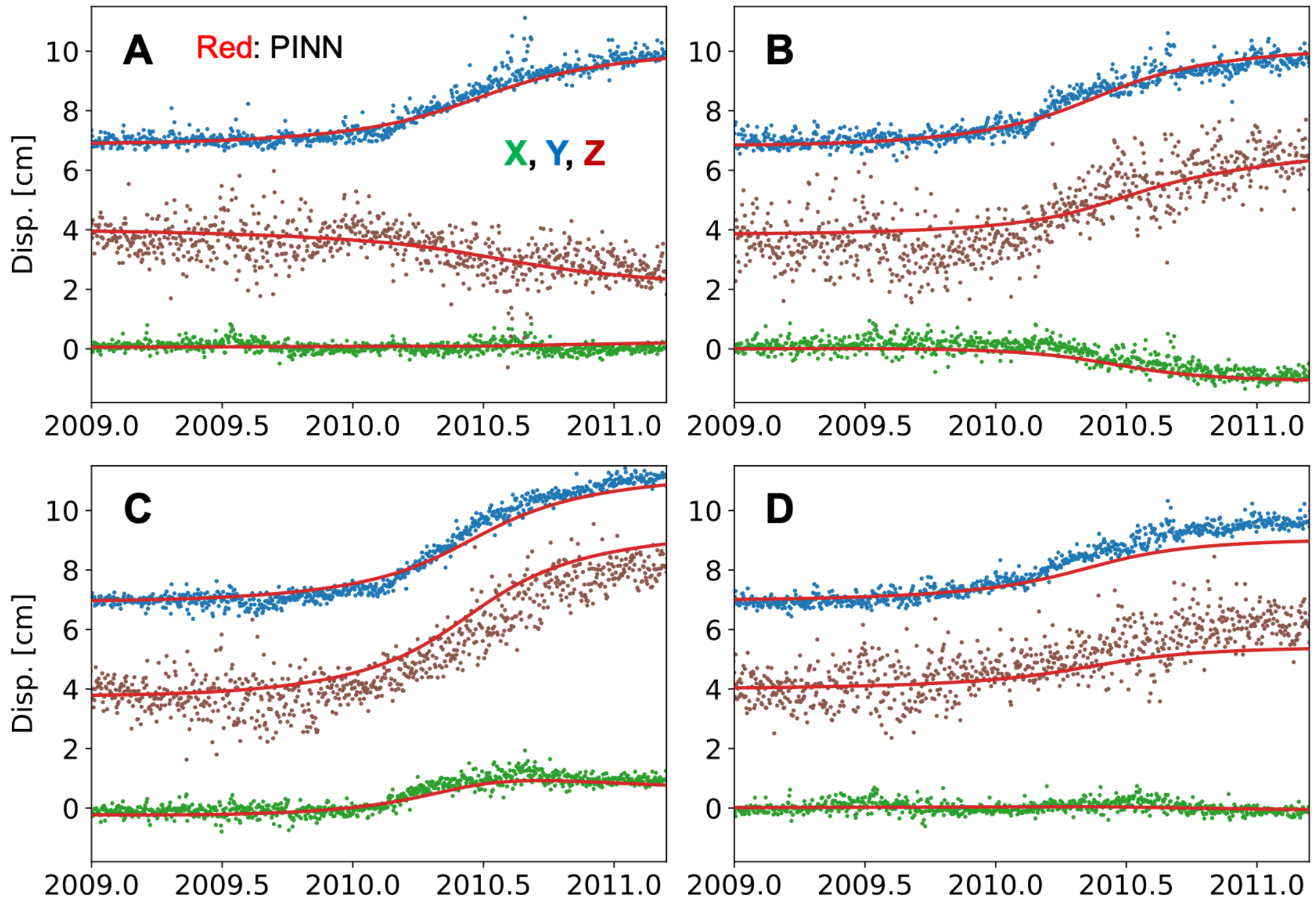

**Extended Data Figure 2.** Expanded time series of surface displacements during 2009-2011 observed at 4 stations. Color dots represent the observation data. Red lines indicate the calculated displacements for PINN-based inversion.

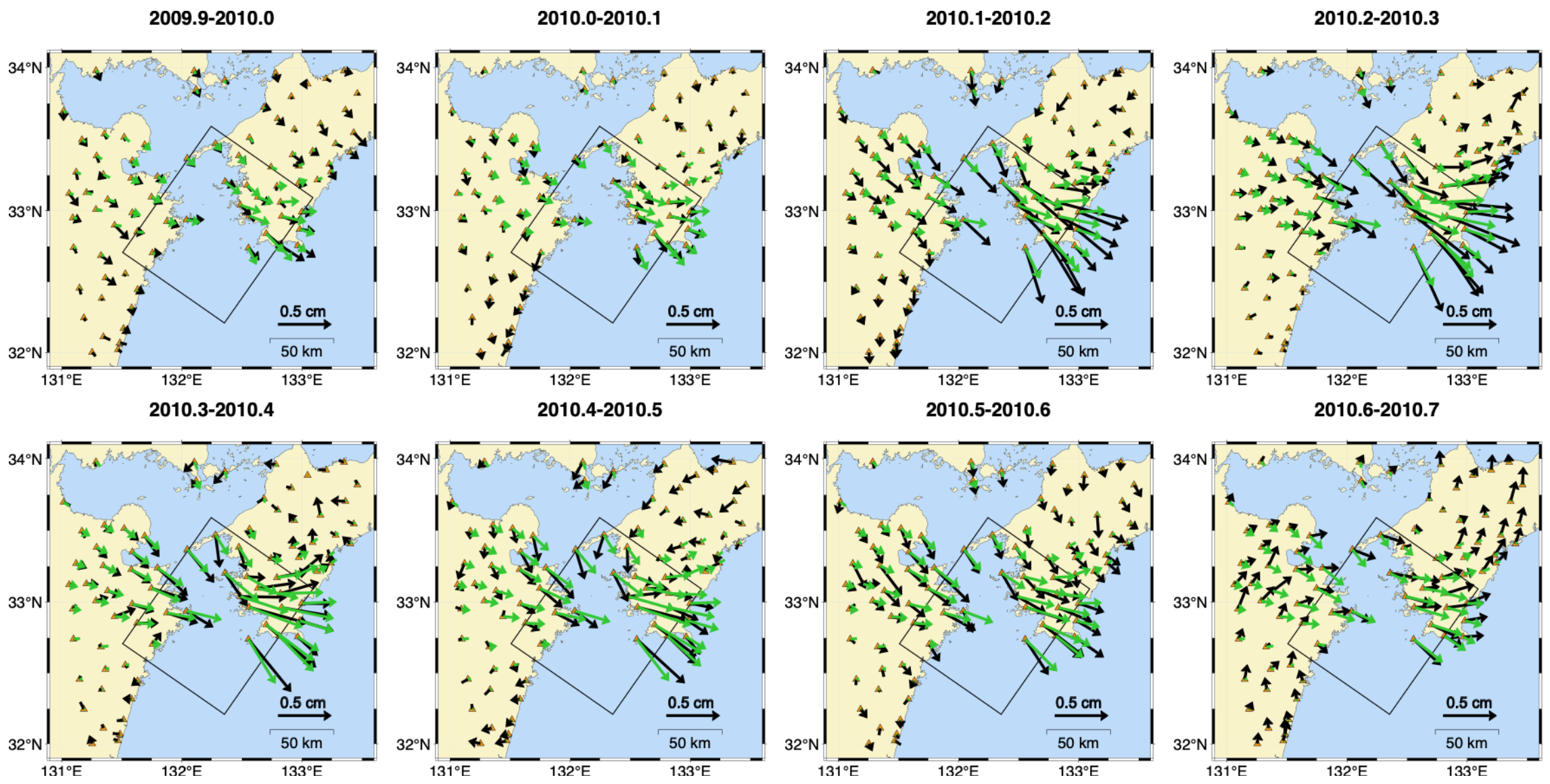

**Extended Data Figure 3.** Displacement field during 2009-2011. Black and green arrows represent the observation and PINN estimation, respectively.

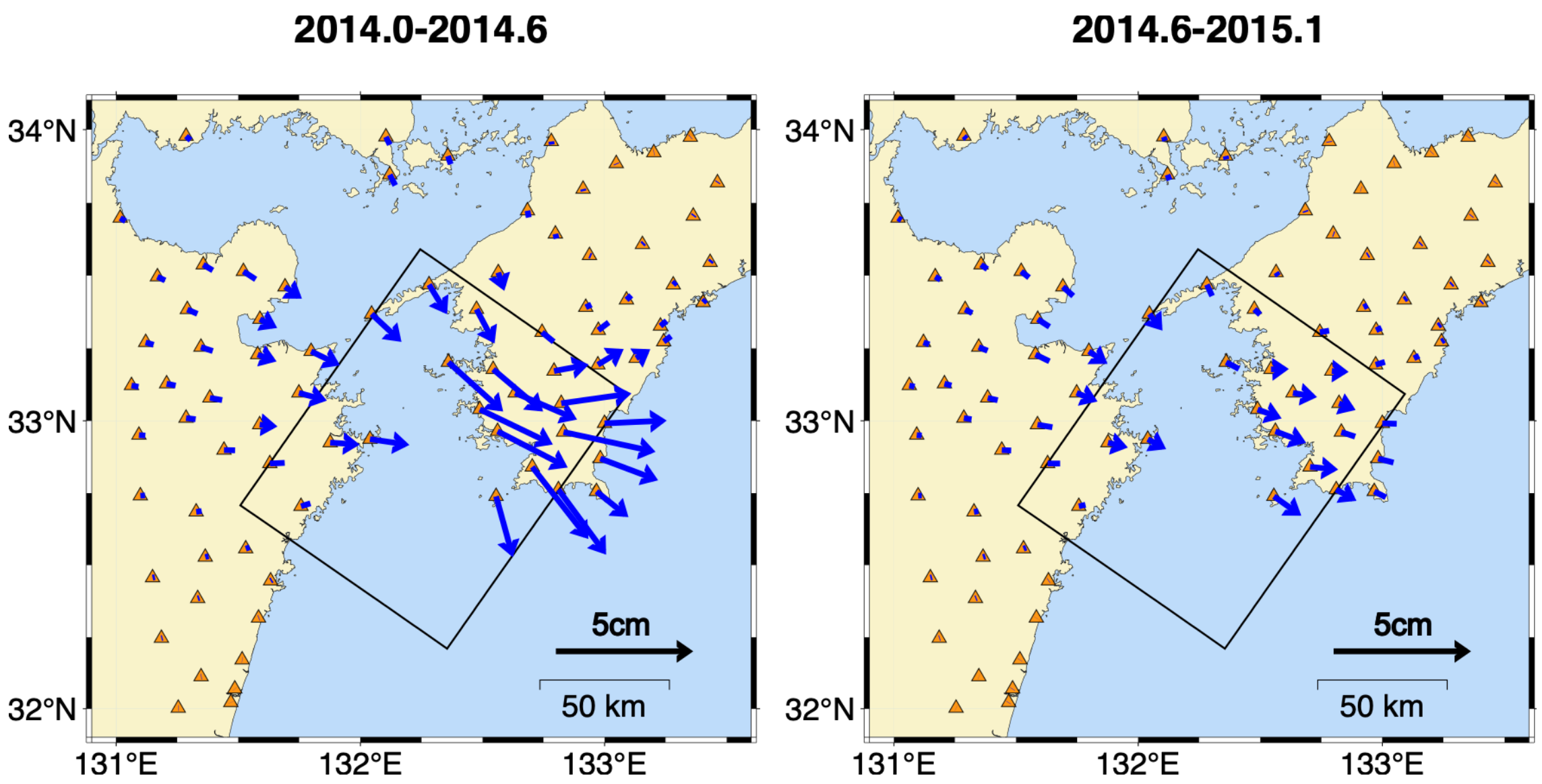

**Extended Data Figure 4.** Predicted displacement field during 2014.0- 2014.6 (left panel) and 2014.6-2015.1 (right panel).

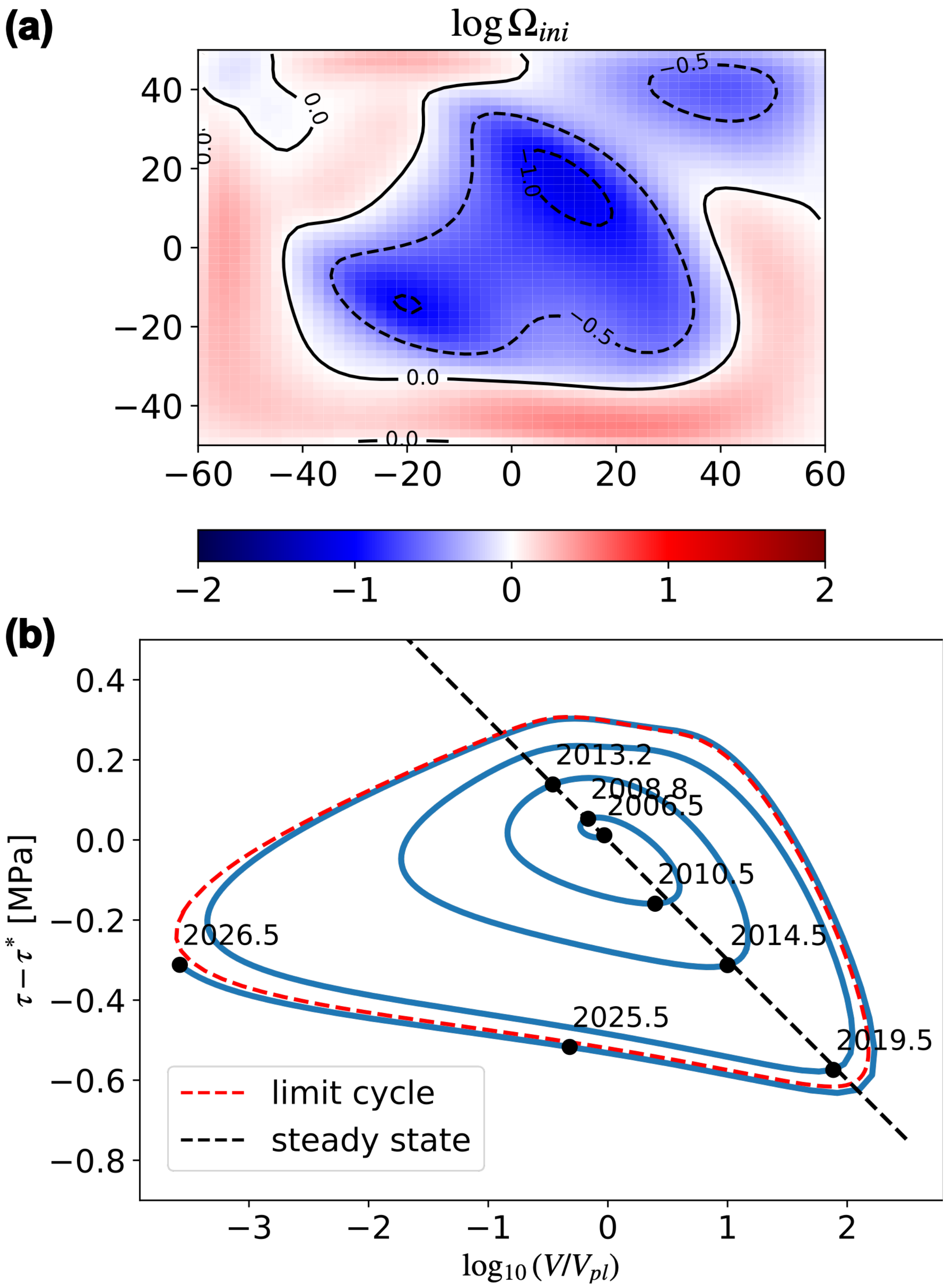

**Extended Data Figure 5.** (a) Fault state distance from steady state: log $\Omega = (\tau - \tau_{ss}) / \sigma b$ at $t$ = 2006.5. (b) $\tau$-$V$ phase plot at the point of minimum nucleation length $R_c$, indicated

by the blue star in Fig.4. The blue line represents the trajectory calculated from the PINN estimated result during 2006.5- 2026.5, and the red line represents the limit cycle trajectory calculated from estimated frictional parameters. The black dashed line shows the steady state stress: $\tau_{ss}(V) - \tau^* = \sigma(a\text{-}b)\ \log(V/V_{pl})$ (Eq. 3).

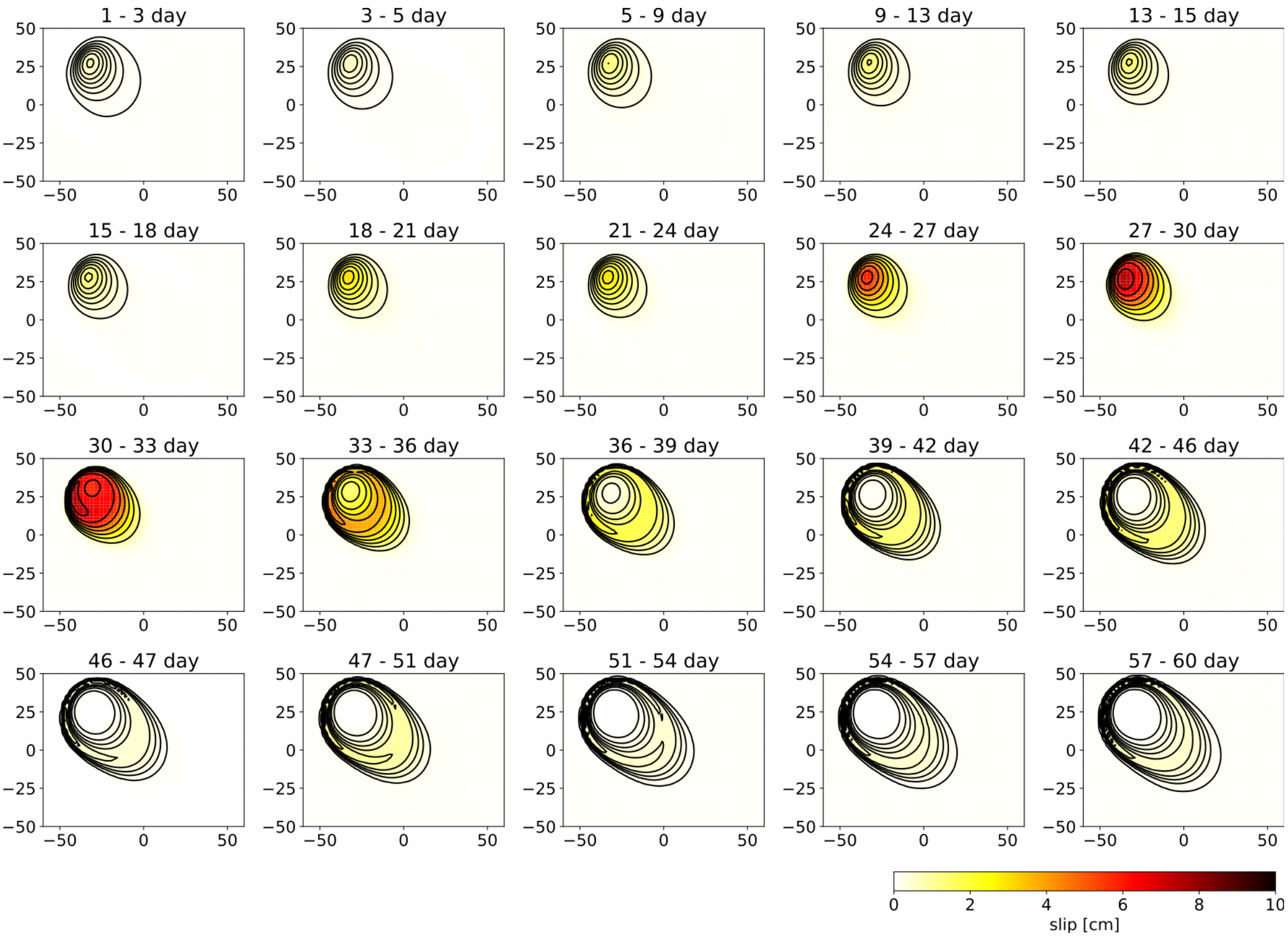

**Extended Data Figure 6.** Slip evolution of the limit cycle event with the estimated frictional parameters. The maximum slip velocity is ~162 $V_{pl}$, and the moment magnitude of this event is 6.8.

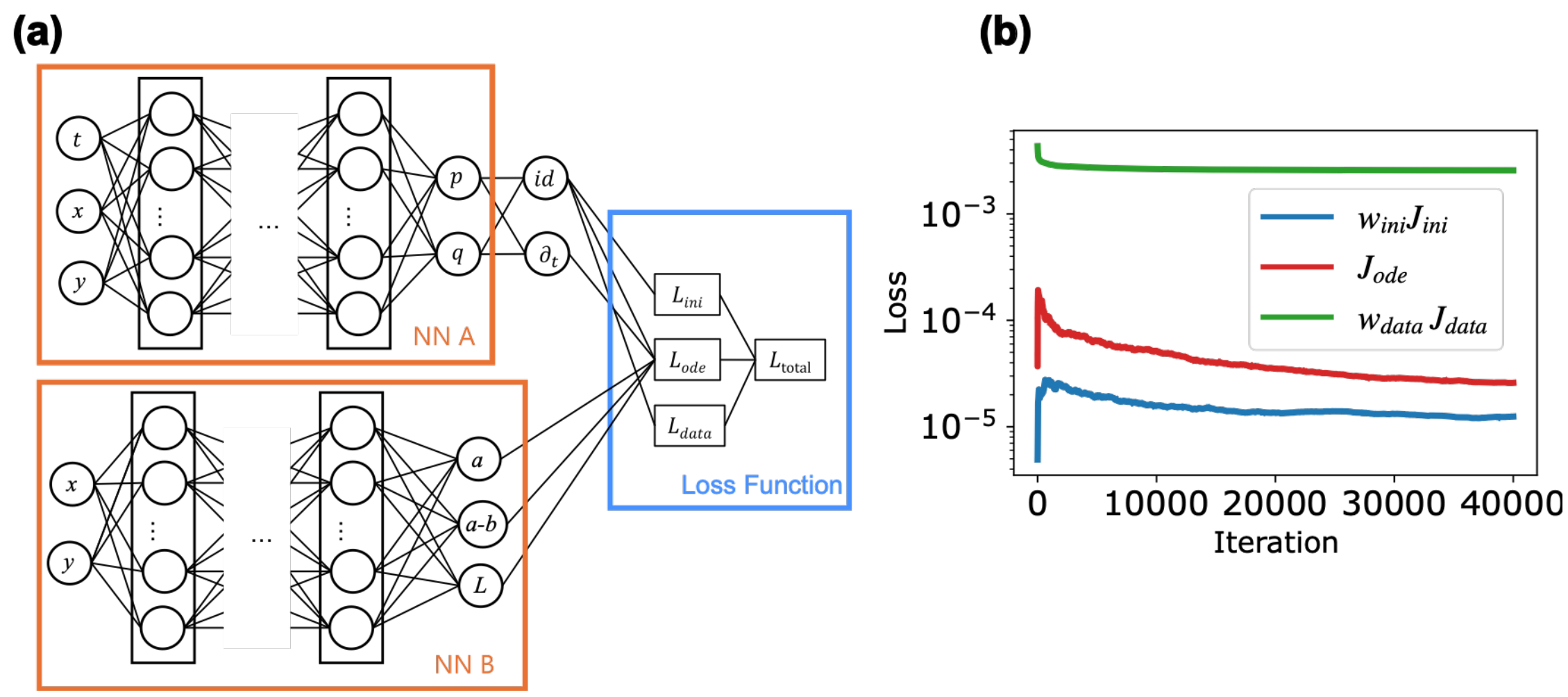

**Extended Data Figure 7.** (a) PINN architecture. The outputs of neural networks A and B represent the simulation variables: p = log($V/V^*$) and q = log($\theta/\theta^*$), and the frictional parameters, respectively. (b) The learning curve of the loss function. Blue, red, and green lines indicate the loss values of $w_{\mathrm{ini}}\,J_{\mathrm{ini}}$, $J_{\mathrm{ode}}$, and $w_{\mathrm{data}}\,J_{\mathrm{data}}$, respectively. After the 40,000 iterations by the L-BFGS method, the loss function converges to $J_{\mathrm{total}} = 2.61\times10^{-3}$, $J_{\mathrm{ini}} = 1.25\times10^{-1}$, $J_{\mathrm{ode}} = 2.59\times10^{-5}$, and $J_{\mathrm{data}} = 2.57$.

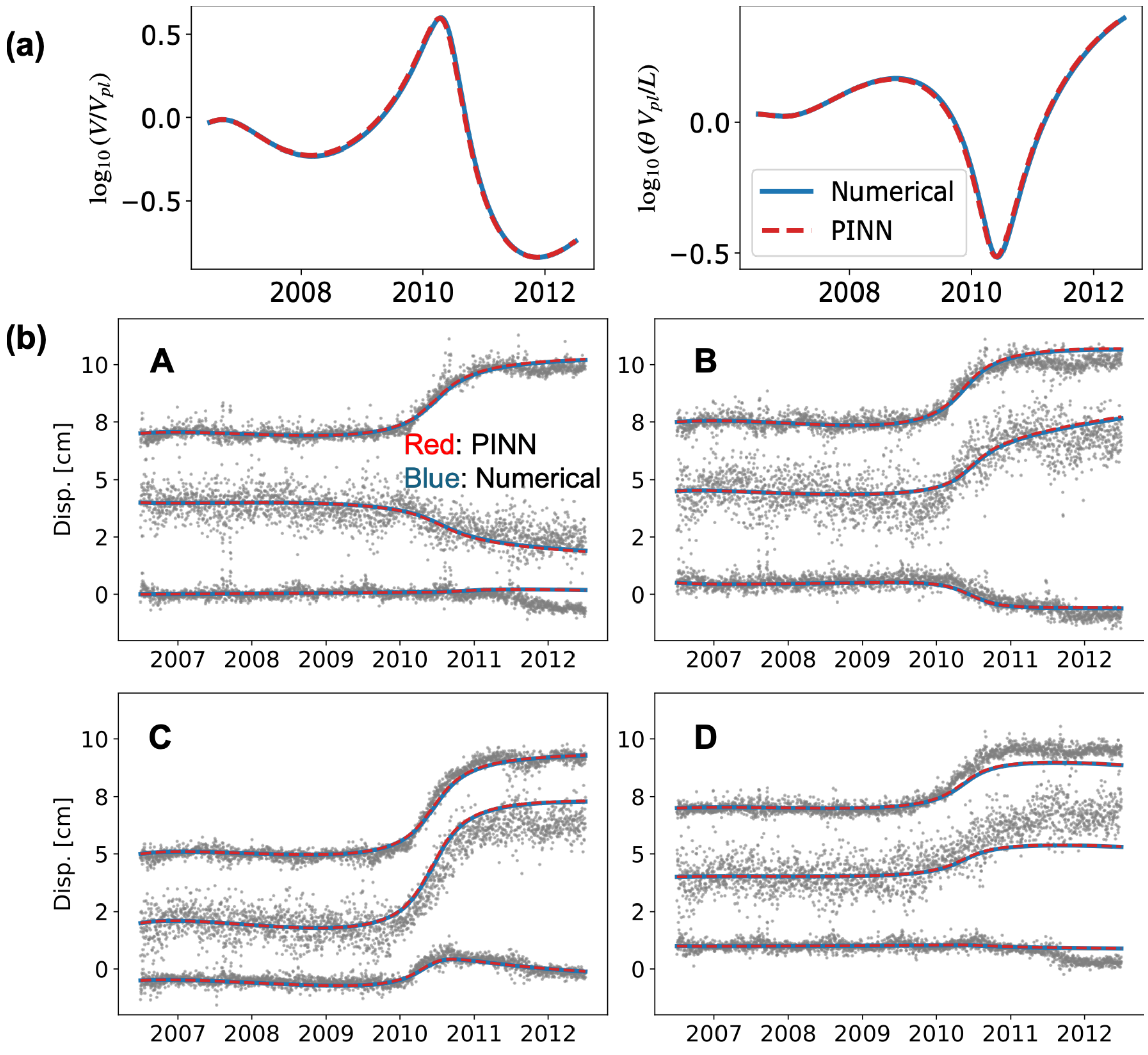

**Extended Data Figure 8.** Comparison between the PINN output and the Runge-Kutta solution with PINN estimated $a$, $a$-$b$, $L$, $V_{\text{ini}}$, and $\theta_{\text{ini}}$. (a) The temporal evolution of $V$ and $\theta$ at the point of minimum nucleation length $R_c$, indicated by the blue star in Fig.4. Red and blue lines represent the PINN output and the Runge-Kutta solution, respectively. (b) Same with Fig.1(b), but the Runge-Kutta solution is plotted as a blue line in addition to PINN output. For better visualization, observation points are plotted as gray dots.

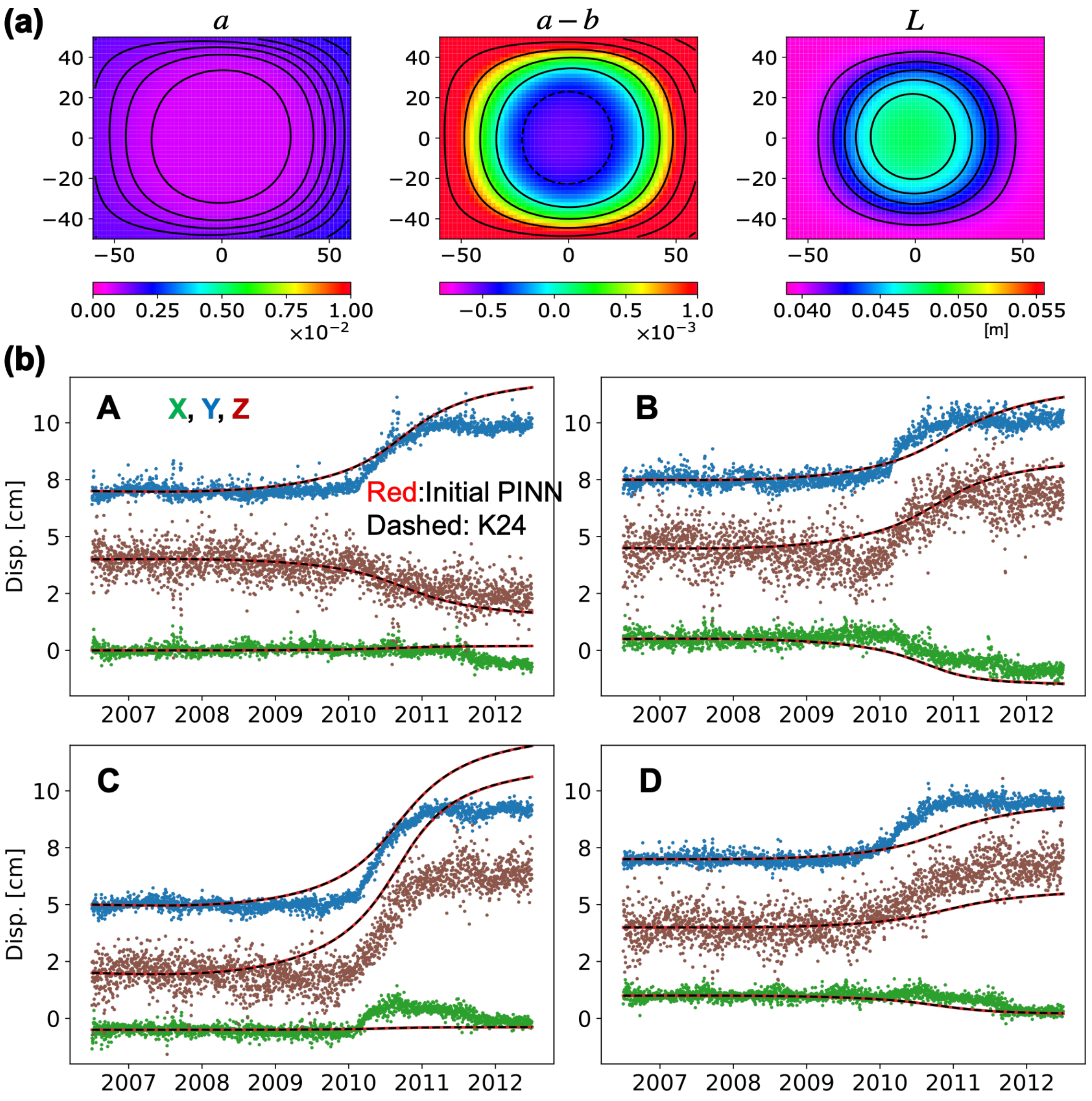

**Extended Data Figure 9.** Initial neural network parameters determined by pre-training with synthetic observation data generated from the K24 estimation model. (a) The spatial distribution of frictional parameters. The velocity weakening region spread out as a circular patch. (b) Same with Fig.1, but for the PINN output at iteration 0.

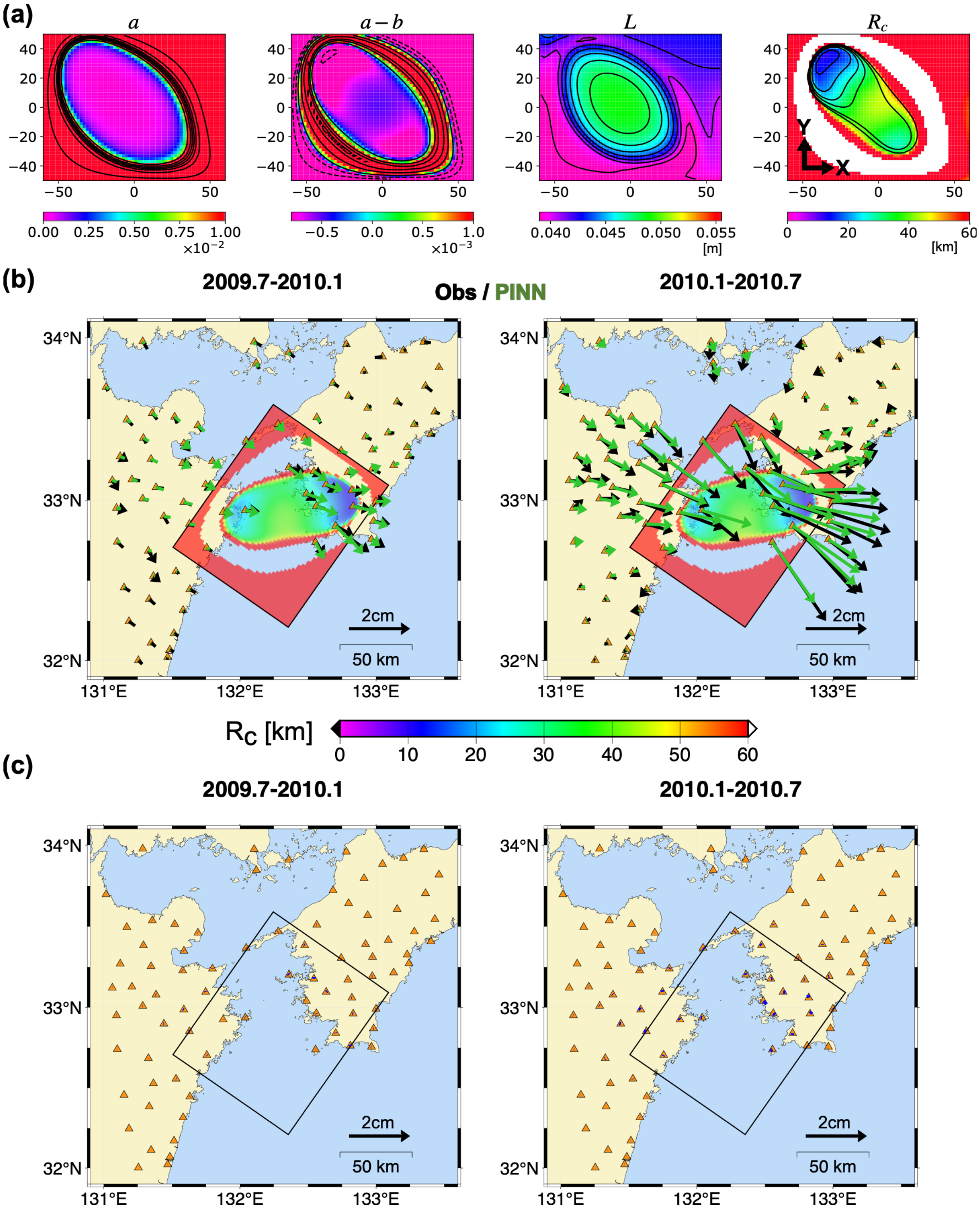

**Extended Data Figure 10.** Results of the PINN-based estimation without the constraints for frictional parameters *a* and *b*. (a) Spatial distribution of estimated frictional parameters for *a*, *a*-*b*, and *L*. $R_c$ shows the nucleation length at the velocity weakening region calculated from the estimated parameters. (b) Same with Fig. 2, but for the PINN

estimation without friction constraints. (c) The differential surface displacement vector between PINN-estimation with and without friction constraints (i.e., the difference between the lower panel of Fig. 2 and Extended Data Fig. 10b). The differential vectors are close to zero on all stations, which means that both estimation results reproduce the same displacement field.